\documentclass[a4paper]{jpconf}
\usepackage{color}
\usepackage[usenames,dvipsnames]{xcolor}
\usepackage{graphicx}
\usepackage{fancyvrb}
\usepackage{amsthm} 
\newcommand*{\fvtextcolor}[2]{\textcolor{#1}{#2}} 
\theoremstyle{definition}
\newtheorem{listing}{Listing}

\usepackage{hyperref}
\hypersetup{colorlinks=true,urlcolor=blue,linkcolor=black}

\begin{document}
\title{GooFit: A library for massively parallelising maximum-likelihood fits}

\author{R~Andreassen$^1$, B~T~Meadows$^1$, M~de~Silva$^1$, M~D~Sokoloff$^1$ and K~Tomko$^2$}
\author{}
\address{$^1$ University of Cincinnati, Physics Department, ML0011, Cincinnati OH 45221-0011, USA}
\address{$^2$ Ohio Supercomputer Center, 1224 Kinnear Road, Columbus OH 43212, USA}
\ead{rolfa@slac.stanford.edu}

\begin{abstract}
Fitting complicated models to large datasets is a bottleneck of
many analyses. We present GooFit, a library and tool for constructing
arbitrarily-complex probability density functions (PDFs) to be evaluated
on nVidia GPUs or on multicore CPUs using OpenMP. The massive parallelisation of dividing up event calculations 
between hundreds of processors can achieve speedups of factors 200-300 in
real-world problems. 
\end{abstract}

\section{Introduction}
\label{sec:intro}

Parameter estimation is a crucial part of many physics analyses. 
GooFit is an interface between the MINUIT minimisation algorithm
and a parallel processor - either a Graphics Processing Unit (GPU) or a multicore CPU - which allows a probability
density function (PDF) to be evaluated in parallel.
Since PDF evaluation on large datasets
is usually the bottleneck in the MINUIT algorithm, this can result
in speedups of up to $\sim 300$ in real problems - which can be the
difference between waiting overnight for the answer, or making a cup
of tea. 

GooFit is named in homage to the fitting package RooFit~\cite{roofit},
with the G standing for GPU.
The overall flow of a 
program using GooFit - create variables, create PDF objects, load data,
fit, as shown in listing~\ref{list:expexp} - should be somewhat familiar to RooFit users.

\section{GooFit code}
\label{sec:usercode}

The original intention of GooFit was to give users access to the parallelising power of CUDA~\cite{cuda-bpg}, nVidia's
programming language for GPUs, 
without requiring them to write CUDA code. By abstracting the thread-management code
using the Thrust library~\cite{thrust}, we have extended the possible parallel platforms;
currently GooFit supports CUDA and OpenMP. In the future we hope also to include
Thrust's TBB backend, as well as any other backends that the Thrust developers add. 

At the most basic level, GooFit objects
representing PDFs, \texttt{GooPdf}s, can be created and combined in plain C++. Only
if a user needs to represent a function outside the existing GooFit classes does she
need to do any CUDA coding; Section~\ref{sec:newpdfs} shows how to create new PDF classes.
We intend, however, that this should be a rarity, and that the existing PDF classes
should cover the most common cases. 

A GooFit program has four main components: 
\begin{itemize}
\item The PDF that models the physical process, represented by a \texttt{GooPdf} object.
\item The fit parameters with respect to which the likelihood is maximised,
represented by \texttt{Variable}s contained in the \texttt{GooPdf}. 
\item The data, gathered into a \texttt{DataSet} object containing one or more \texttt{Variable}s.
\item A \texttt{FitManager} object which forms the interface between Minuit 
(or, in principle, any maximising algorithm) and the \texttt{GooPdf}.
\end{itemize}
Listing~\ref{list:expexp} shows a simple fit of an exponential function.

\begin{listing}
\label{list:expexp} 
\emph{Fit for unknown parameter $\alpha$ in $e^{\alpha x}$. GooFit classes
are shown in red, important operations in blue.}

\begin{Verbatim}[commandchars=\\\$\#]
int main (int argc, char** argv) {
  \fvtextcolor$violet#$// Independent variable (name, lower limit, upper limit)#
  \fvtextcolor$red#$Variable*# xvar = new \fvtextcolor$red#$Variable#("xvar", 0, log(1+RAND_MAX/2)); 
  
  \fvtextcolor$violet#$// Create data set#
  \fvtextcolor$red#$UnbinnedDataSet# data(xvar);
  for (int i = 0; i < 100000; ++i) {
    \fvtextcolor$violet#$// Generate toy event#
    xvar->value = xvar->upperlimit - log(1+rand()/2);
    \fvtextcolor$violet#$// ...and add to data set.# 
    \fvtextcolor$blue#$data.addEvent();# 
  }
  
  \fvtextcolor$violet#$// Create fit parameter (name, initial value, step size, lower and upper limit)# 
  \fvtextcolor$red#$Variable*# alpha = new \fvtextcolor$red#$Variable#("alpha", -2, 0.1, -10, 10);
  \fvtextcolor$violet#$// Create GooPdf object - name, independent variable, fit parameter.#
  \fvtextcolor$red#$ExpPdf*# exppdf = new \fvtextcolor$red#$ExpPdf#("exppdf", xvar, alpha); 
  \fvtextcolor$violet#$// Move data to GPU#
  \fvtextcolor$blue#$exppdf->setData(&data);#

  \fvtextcolor$red#$FitManager# fitter(exppdf);
  \fvtextcolor$blue#$fitter.fit();# 

  return 0;
}
\end{Verbatim}
\end{listing}

The example code contains two different uses of the \texttt{Variable} class:
\texttt{xvar} represents the measured (in this toy example, randomly generated)
experimental results, while \texttt{alpha} represents the model parameter
to be determined - in this case a decay constant. In the latter case we supply
an initial value for the fit to start with and a guess at a reasonable step size,
in addition to upper and lower allowed limits. For data points, which will not vary in 
the fit, the initial value and step size are not needed, so we use a \texttt{Variable}
constructor which does not require them. 

The \texttt{UnbinnedDataSet} class is supplied, at construction time, with  
pointers to the \texttt{Variable}s it is to contain - in general by means of
a \texttt{vector} of \texttt{Variable*}, but for convenience in the single-observable
case there is a constructor that takes a single pointer. It is then filled by 
means of the \texttt{addEvent} method, which creates a ``row'' or ``event'' within
the dataset, containing the values, at the time of the \texttt{setData} call, of
the comprising \texttt{Variable}s. An example may be helpful to visualising this:
\begin{Verbatim}[commandchars=\\\$\#]
  \fvtextcolor$red#$Variable*# xvar = new \fvtextcolor$red#$Variable#("xvar", 0, log(1+RAND_MAX/2)); 
  \fvtextcolor$red#$UnbinnedDataSet# data(xvar); \fvtextcolor$violet#$ // Data set is empty#
  xvar->value = 3; \fvtextcolor$violet#$ // Still empty#
  data.addEvent(); \fvtextcolor$violet#$ // Now contains one event, value 3#
  xvar->value = 5; 
  data.addEvent(); \fvtextcolor$violet#$ // Two events: 3, 5#
  xvar->value = 1; 
  data.addEvent(); \fvtextcolor$violet#$ // Three events: 3, 5, 1#
\end{Verbatim}
Note that \texttt{UnbinnedDataSet} stores its events in host memory,
that is, not on the GPU; not until the \texttt{setData} method of a
\texttt{GooPdf} is called are the events moved to the GPU or other
target device. 

\begin{figure}
\begin{center}
\includegraphics[width=0.48\textwidth]{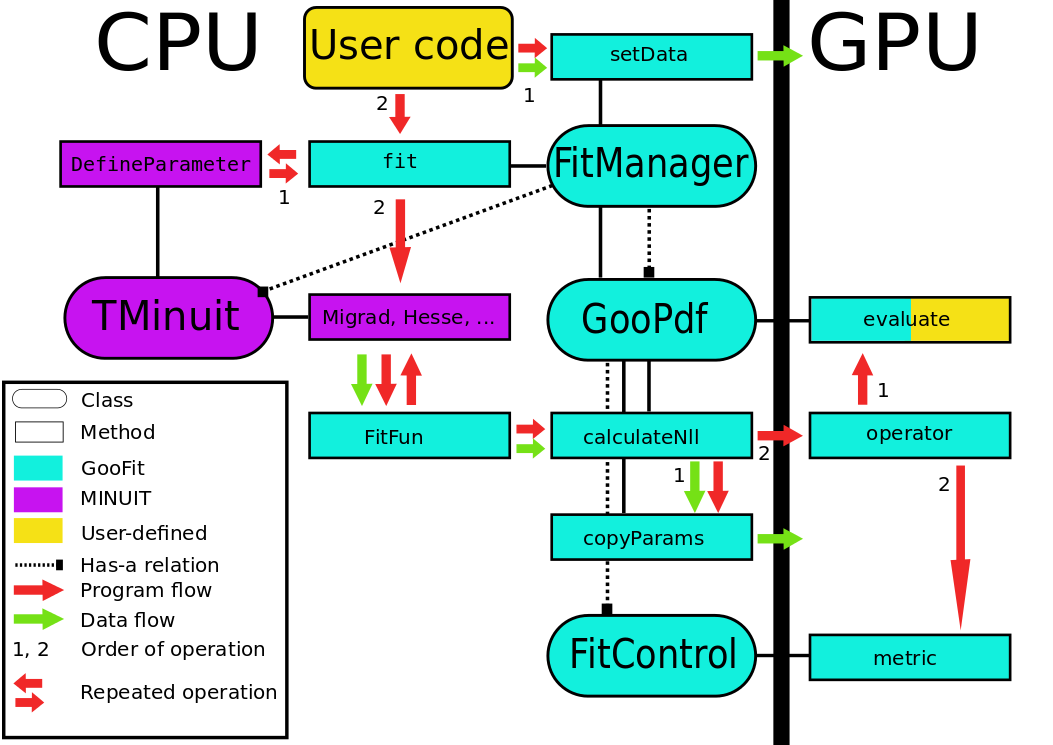}
\includegraphics[width=0.48\textwidth]{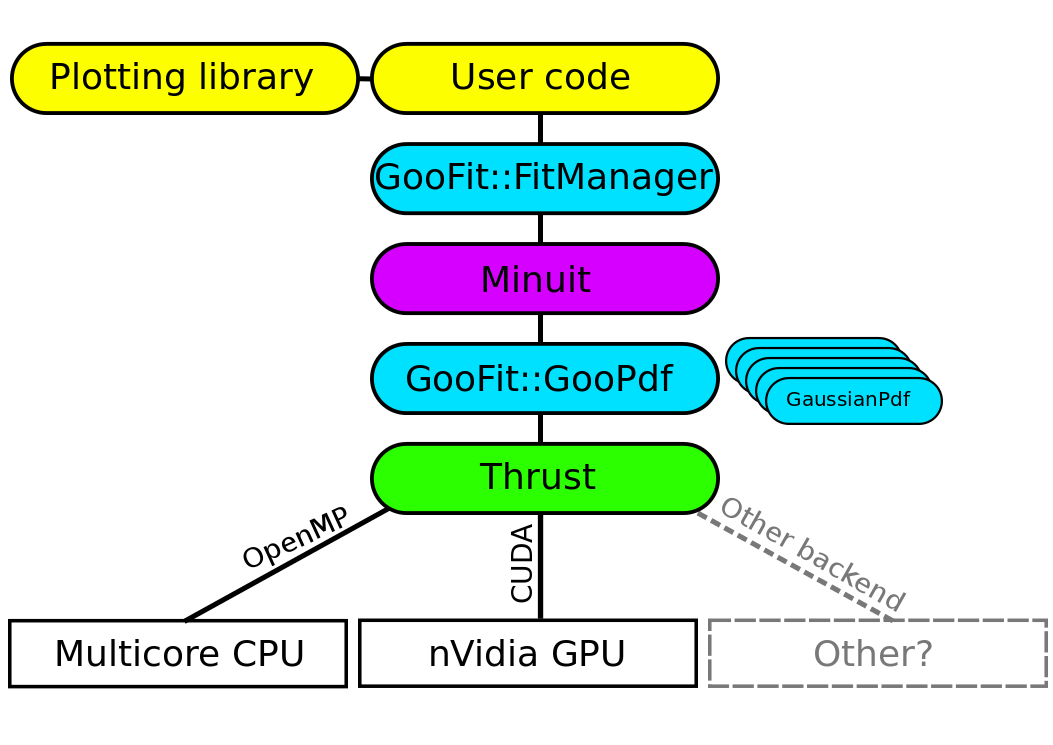}
\end{center} 
\caption{\label{fig:goofitflow} Control and data flow of a GooFit program (left)
and overall architecture of GooFit (right).} 
\end{figure}

The \texttt{fit} method of the \texttt{FitManager} class does two
things: First, it sets up the \texttt{Minuit}~\cite{jam72,davi91} fit by calling the
\texttt{DefineParameter} method on each \texttt{Variable} representing
a fit parameter. Second, it passes control to MINUIT's \texttt{mnmigr}
method, which will in turn call GooFit's evaluation method for different
sets of parameters until it converges or gives up. The program flow
is illustrated in Figure~\ref{fig:goofitflow}. Notice, on the device side, 
the separate calls to the \texttt{evaluate} method under \texttt{GooPdf}, 
and the \texttt{metric} method under \texttt{FitManager}; this allows switching
between, for example, maximum-likelihood and chi-squared fits without
changing any PDF creation code. 

Calls to the device-side methods are done by the Thrust template
library; this allows us to switch the execution backend by means of
a compile-time switch. In particular, GooFit currently supports CUDA
and OpenMP parallelisation. Figure~\ref{fig:goofitflow} shows the
overall organisation of GooFit, with user code at the top and Thrust
talking to the parallelising target at the bottom. Note that Minuit,
in the middle, can in principle be replaced by any algorithm for searching
through parameter space.

\section{Combining functions}
\label{sec:combining}

To create arbitrarily complex PDFs, the user can combine 
simple ones like the exponential and Gaussian in several
ways, the most common being addition and multiplication.
For example, a two-dimensional distribution, described by
a separate exponential in each of two variables $x$ and $y$, may be represented
as a \texttt{ProdPdf} object, as shown in Listing~\ref{list:prodpdf}.

\begin{listing}
\label{list:prodpdf} \emph{Product of two exponentials.}

\begin{Verbatim}[commandchars=\\\$\#]
int main (int argc, char** argv) {
  Variable* xvar = new Variable("xvar", 0, log(1+RAND_MAX/2)); 
  Variable* yvar = new Variable("yvar", 0, log(1+RAND_MAX/2)); 
  
  vector<Variable*> varList;
  varList.push_back(xvar);
  varList.push_back(yvar);
  UnbinnedDataSet data(varList);
  for (int i = 0; i < 100000; ++i) {
    xvar->value = xvar->upperlimit - log(1+rand()/2);
    yvar->value = yvar->upperlimit - log(1+rand()/2);
    data.addEvent(); 
  }
  
  Variable* alpha_x = new Variable("alpha_x", -2.4, 0.1, -10, 10);
  Variable* alpha_y = new Variable("alpha_y", -1.1, 0.1, -10, 10);
  vector<\fvtextcolor$red#$PdfBase*#> pdfList;
  pdfList.push_back(new ExpPdf("exp_x", xvar, alpha_x)); 
  pdfList.push_back(new ExpPdf("exp_y", yvar, alpha_y)); 

  \fvtextcolor$red#$ProdPdf*# product = new \fvtextcolor$red#$ProdPdf#("product", pdfList);
  product->setData(&data);

  FitManager fitter(product);
  fitter.fit();

  return 0;
}
\end{Verbatim}
\end{listing}

In addition to products, GooFit implements sums of PDF through
the \texttt{AddPdf} class, convolutions by means of the \texttt{ConvolutionPdf},
and function composition, ie $f(g(x))$, in the \texttt{CompositePdf}
class. Finally, the \texttt{MappedPdf} allows the construction of functions
with different form depending on a set parameter, that is 
\begin{eqnarray*} 
F(x) &=& \left\{ \begin{array}{rl} 
A(x) & \mbox{  if $x \in [x_0, x_1)$}\\
B(x) & \mbox{  if $x \in [x_1, x_2)$}\\
\ldots & \\ 
Z(x) & \mbox{  if $x \in [x_{N-1}, x_N]$}
\end{array}\right. 
\end{eqnarray*}

All these combined PDF classes can be nested arbitrarily deeply. 

\section{Adding new PDFs}
\label{sec:newpdfs}

For advanced users who need additional PDFs, GooFit makes
it easy to write a new PDF class. There are two steps to the process:
\begin{itemize}
\item Write an evaluation method with the required signature,
using the provided index array to look up parameters and observables.
An example is shown in Listing~\ref{list:evalexample}.
\item Create a C++ class inheriting from \texttt{GooPdf}, in which
the constructor populates the index array through the \texttt{registerParameter}
method. Listing~\ref{list:indexexample} shows an example. 
\end{itemize}
That's it! Putting the new \texttt{FooPdf.cu} and \texttt{FooPdf.hh} files
in the \texttt{PDFs} directory of GooFit will cause them to be compiled with
the rest of the framework, and be available for use in the same way as the
pre-existing PDFs. 

\begin{listing}
\label{list:evalexample} \emph{Evaluation function for a Gaussian
PDF. Note the double indirection of the index-array lookups.
Here \texttt{fptype} indicates a floating-point number, by default \texttt{double}
precision.}

\begin{Verbatim}[commandchars=\\\$\#]
__device__ fptype device_Gaussian (fptype* evt, fptype* p, unsigned int* indices) {
  fptype x = evt[indices[2 + indices[0]]]; 
  fptype mean = p[indices[1]];
  fptype sigma = p[indices[2]];

  fptype ret = EXP(-0.5*(x-mean)*(x-mean)/(sigma*sigma));
  return ret; 
}

__device__ device_function_ptr ptr_to_Gaussian = device_Gaussian; 
\end{Verbatim}
\end{listing}

\begin{listing}
\label{list:indexexample} \emph{Populating the index array used in Listing~\ref{list:evalexample}}.

\begin{Verbatim}[commandchars=\\\$\#]
__host__ GaussianPdf::GaussianPdf (std::string n, Variable* _x, 
                                   Variable* mean, Variable* sigma) 
  : GooPdf(_x, n) 
{
  std::vector<unsigned int> pindices;
  pindices.push_back(registerParameter(mean));
  pindices.push_back(registerParameter(sigma));
  cudaMemcpyFromSymbol((void**) &host_fcn_ptr, ptr_to_Gaussian, sizeof(void*));
  initialise(pindices); 
}

\end{Verbatim}
\end{listing}

\section{Results}
\label{sec:tddp}

To ensure the usefulness of GooFit in real-world physics problems
we have, simultaneously with developing GooFit, used it as our fitting
tool in a time-dependent Dalitz-plot analysis of the decay $D^0\to\pi^+\pi^-\pi^0$. 
This fit has been our ``driver'' for GooFit, in that every
time we needed a feature for the physics, we added it to the GooFit
engine. 
This mixing fit is rather complex, involving, for the signal component, 
16 amplitudes (each a complex number) to describe the Dalitz-plot distribution,
a time component where hyperbolic and trigonometric functions
are convolved with Gaussian resolution functions, and a distribution
of the uncertainty $\sigma_t$ on the decay time which varies across
the Dalitz plot. All in all, there are about 40 free parameters in the
fit, and the data set is roughly a hundred thousand events; running this
on one core of a modern CPU takes about five hours, depending on the 
data. Using GooFit with a CUDA backend, this is reduced to a much more comfortable one minute,
a speedup factor in the region of 300, relative to the original hand-coded
CPU implementation. Testing with the OpenMP backend, as shown in
Table~\ref{tab:speedups}, indicates that some part of this speedup
is due to the reorganising of the code, or perhaps of the memory, whose
management differs between the original CPU implementation and GooFit. 

In addition to the mixing fit described above, we have tested GooFit
on ``Zach's fit'', named for a (now graduated) student of our group. 
This is a binned fit, extracting the $D^{*+}$ line width from measurements
of the $D^{*+}-D$ mass difference, and involving an underlying relativistic P-wave Breit-Wigner
convolved with a resolution function comprising several Gaussians. In its
original RooFit implementation this fit takes about 7 minutes on our
workstation `Cerberus'. With GooFit this is reduced down to a few seconds;
however, the speedup is not as impressive as with the mixing fit. We do
not fully understand the differences, but believe it is partly because
a binned fit, with only a few thousand PDF evaluations per MINUIT iteration, 
does not take as much advantage of the massive parallelisation of the GPU
as does an unbinned fit with a hundred thousand events. 

As shown in Table~\ref{tab:platforms}, we have measured GooFit's performance on three different
platforms: Our workstation Cerberus, the `Oakley' computer farm of the Ohio Supercomputer
Center, and the laptop `Starscream' with a mid-range 650M GPU. In addition
we have tested earlier versions of GooFit on K10 and K20 boards,
with speedups of about a factor 2 relative to the C2050. 

The execution times shown in Table~\ref{tab:speedups} are for Minuit fits using the Migrad
algorithm; time to load data into memory and create PDF objects is not shown,
on the assumption that these are negligible for realistic problems. It is worth
noting that in going from the original implementations of the two fits (run on one
core of Cerberus) to the GooFit implementation with one OpenMP thread, there is
already a speedup in the range of 6-7. We believe that this is due to differences
in memory layout - GooFit represents its events as a single large array of floating-point
numbers - and, in the case of the Zach fit, perhaps also to the large number 
of virtual-function calls in inner loops of the RooFit implementation. 

As we increase the number of OpenMP threads, we see a nearly-linear speedup, ie times
inversely proportional to the number of threads, up until the number of threads equals
the number of cores. After this there is no gain from adding more threads, except
for the sweet spot of having exactly twice as many threads as there are physical cores - 
this takes full advantage of the hyperthreading capacity of these processors, and gives
a 30\% speedup on Cerberus and a 50\% speedup on the more recent Starscream. 
It is clear that the power of a GPU to speed up fits varies with the precise
problem, but in the best case a C2050 GPU can be 5 times as fast as the same
code parallelised using OpenMP and taking full advantage of a powerful dual quad-core
CPU. 

\begin{table}
\begin{center}
\begin{tabular}{|l|ccccc|}
\hline
Name             &        Chip      &   Cores        & Clock [GHz]  & RAM [Gb] & OS \\ \hline
Cerberus (CPU)   & Intel Xeon E5520 &      8*        &   2.27             &  24      &  Fedora 14       \\
Cerberus (GPU)   & nVidia C2050     &     448        &   1.15             &   3      &  Fedora 14       \\ \hline
Starscream (CPU) & Intel i7-3610QM  &      4*        &   2.3              &   8      &  Ubuntu 12.04    \\
Starscream (GPU) & nVidia 650M      &     384        &   0.9              &   1      &  Ubuntu 12.04    \\ \hline
Oakley           & nVidia C2070     &     448        &   1.15             &   6      &  RedHat 6.3      \\ \hline
\end{tabular}
\end{center}
\caption{\label{tab:platforms} Some specifications of the testing platforms.
Asterisks next to the number of cores indicate hyperthreading - two virtual processors per physical core. 
}
\end{table}

\begin{table}
\begin{center}
\begin{tabular}{|l|cc|cc|}\hline  
& \multicolumn{2}{|c|}{Mixing fit} 
& \multicolumn{2}{|c|}{Zach's fit} \\ \hline
Platform           & Time [s] & Speedup & Time [s] & Speedup   \\ \hline
Original CPU       &  19489   &   1.0   &   438      &  1.0    \\ \hline
Cerberus OMP (1)   &   3056   &   6.4   &    60.6    &  7.2    \\
Cerberus OMP (2)   &   1563   &  12.5   &    31.0    & 14.1    \\
Cerberus OMP (4)   &    809   &  24.1   &    18.2    & 24.1    \\
Cerberus OMP (8)   &    432   &  45.1   &     9.2    & 47.6    \\
Cerberus OMP (12)  &    534   &  36.5   &    12.2    & 35.9    \\
Cerberus OMP (16)  &    326   &  59.8   &     6.9    & 63.5    \\
Cerberus OMP (24)  &    432   &  45.1   &     9.5    & 46.1    \\
Cerberus C2050     &     64   & 304.5   &     5.8    & 75.5    \\
 \hline
Starscream OMP (1) &   2042   &   9.5   &   37.1   &  11.8   \\
Starscream OMP (2) &   1056   &  18.5   &   19.2   &  22.8   \\
Starscream OMP (4) &    562   &  34.6   &   10.8   &  40.6   \\
Starscream OMP (8) &    407   &  47.9   &    6.9   &  63.5   \\
Starscream 650M    &    212   &  91.9   &   18.6   &  23.5   \\
\hline
Oakley C2070       &     54   & 360.1   &    5.4  &   81.1   \\
\hline 
\end{tabular}
\end{center}
\caption{\label{tab:speedups} Timing results and speedups for the test fits.}
\end{table}

\section{Summary}
\label{sec:summary}

We have developed GooFit for use in real-world physics problems, 
and have achieved speedups of $\sim 300$ in particular analyses. 
We have created a robust framework that is easy enough
for a new graduate student to use, but flexible enough that the
most advanced analyses will find it useful. 

GooFit's source code lives in a GitHub repository at \url{https://github.com/GooFit},
which also includes a manual and several example fits to get
the new user started.

\ack 

Development of GooFit is supported by NSF grant NSF-1005530.
We are grateful for valuable suggestions and help from Cristoph
Deil and feedback and user reports from Olli Lupton and Stefanie Reichert.
Jan Veverka and Helge Voss contributed implementations
of the bifurcated Gaussian and Landau distributions. 
The Ohio Supercomputer Center made their computer farm ``Oakley'' available
for development, for testing, and for GooFit outreach
workshops. nVidia's Early Access Program made it possible to
test our code on K10 and K20 boards.

\section*{References} 
\bibliography{pif}{}
\bibliographystyle{plain}

\end{document}